\documentclass[twocolumn,showpacs,preprintnumbers,fleqn,amsmath,amssymb,aps]{revtex4}
\usepackage{graphicx}
\usepackage{dcolumn}
\usepackage{bm}
\usepackage{amsmath}
\usepackage{amssymb}
\usepackage{graphics}

\begin{document}
\title{Relaxation of transport properties in electron doped  ${\rm SrTiO_3}$}
\author{Moty Schultz}
\author{Lior Klein}

\affiliation{Department of Physics, Nano-magnetism Research Center,
Institute of Nanotechnology and Advanced Materials, Bar-Ilan
University, Ramat-Gan 52900, Israel}

\date{\today}


\begin{abstract}

We electron-dope single crystal samples of $\rm SrTiO_3$ by exposing
them to Ar$^+$ irradiation and observe carrier mobility similar in
its magnitude and temperature dependence to the carrier mobility in
other electron-doped $\rm SrTiO_3$ systems.  We find that some
transport properties are time-dependent. In particular, the sheet
resistance increases with time at a temperature-dependent rate,
suggesting an activation barrier on the order of 1 eV. We attribute
the relaxation effects to diffusion of oxygen vacancies - a process
with energy barrier similar to the observed activation energy.

\end{abstract}

\maketitle

Perovskites attract considerable interest for their wide range of
intriguing properties, including colossal magnetoresistance in
manganites \cite{colossal1,colossal2}, high-$T_C$ superconductivity
in cuprates \cite{cuprates}, ferroelectricity in titanates
\cite{ferroelectricity}, and itinerant magnetism (ferromagnetism and
antiferromagnetism) in ruthenates \cite{AFM,FM}. In addition to
their individual intriguing properties, for applications, it is
particularly appealing that perovskites-based heteroepitaxial
structures can be grown epitaxially, commonly on $\rm SrTiO_3$, thus
enabling a wide spectrum of new functionalities which may form the
basis for future oxide electronics.

In addition to serving  as a substrate for perovskite films, $\rm
SrTiO_3$ may be used to produce high mobility conductors that would
be useful in future oxide electronics. A familiar way to obtain a
$\rm SrTiO_3$ - based high mobility conductor is by electron doping
\cite{electron mobility,elec doping}. Recently it has been
demonstrated that ${\rm SrTiO_3-LaAlO_3}$ heterostructures prepared
in a particular way also yield high mobility conductivity. Some
groups attributed this phenomenon to the formation of a quasi two
dimensional electron gas at the ${\rm SrTiO_3-LaAlO_3}$ interface
due to polarity discontinuity \cite{polarity_dis,high mobility},
while others argue that it is related to the formation of oxygen
vacancies \cite{origin perspectives,the role,origin unusual}.

Both methods yield high mobilities on the order of $\rm{ 10,000 \
cm^2V^{-1}s^{-1}}$ at 4.2 K \cite{origin perspectives,electron
mobility,elec doping,polarity_dis,high mobility,the role,origin
unusual}, suggesting ${\rm SrTiO_3}$ may be an important component
in oxide-based electronic devices. Some possibilities for such use
have been demonstrated already in its use as a gate \cite{mott
transition} and a channel \cite{sto based,fet} in field effect
transistors.

Electron doping ${\rm SrTiO_3}$ is commonly achieved by creating
oxygen vacancies which transform  ${\rm SrTiO_3}$ into ${\rm
SrTiO_{3-\delta}}$. Oxygen vacancies may be induced in various ways
including high-temperature annealing in oxygen reduced pressure, and
Ar$^+$-irradiation \cite{electronic transport,elec doping, localized
metallic, electron mobility, surface}. Ar$^+$ irradiation is also
the method that we use to electron-dope our samples.

For any future applications of electron-doped ${\rm SrTiO_3}$, it is
important to elucidate the stability of its electrical properties
over time. For this reason, in this report we focus on relaxation
effects of electrical transport in electron-doped ${\rm SrTiO_3}$.
We have irradiated single crystal samples of SrTiO3 with Ar$^+$ and
explored the changes in the sheet resistance, mobility and
magnetoresistance (MR). We find that while the sheet resistance
changes with time (although qualitatively it remains unchanged), the
mobility and the MR are time-independent. Our analysis indicates
that the activation energy for the observed relaxation is about 1 eV
which is the energy scale observed for diffusion of oxygen
vacancies. This suggests that diffusion of oxygen vacancies is
responsible for the observed relaxation effects.

\begin{figure}
\begin{center}
\includegraphics [scale=0.6]{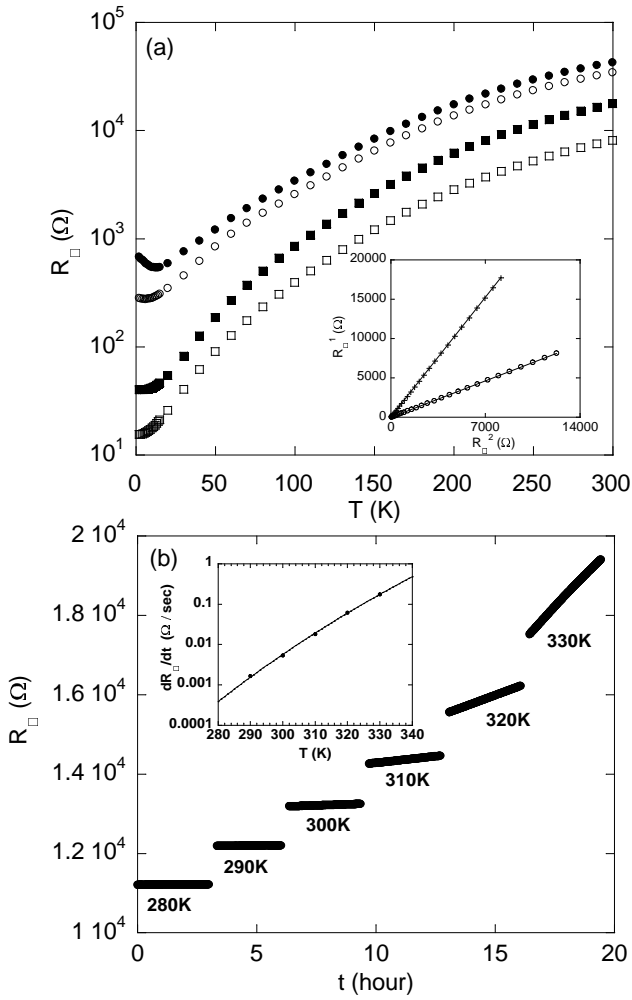}
\end{center}
\caption{(a) Sheet resistance ($R_\Box$) of Ar$^+$  irradiated ${\rm
SrTiO_3}$ as a function of temperature after 30 (squares) and 90
(circles) seconds of irradiation. Open and full symbols are used for
data taken shortly after irradiation and several days after
irradiation, respectively . Inset:  $R_\Box$ after 90 seconds of
irradiation as a function of $R_\Box$ after 60 seconds of
irradiation ($\circ$). $R_\Box$ after 90 seconds of irradiation and
several days of waiting, as a function of $R_\Box$ measured shortly
after the irradiation (+). The lines are fits to a liner function.
(b) $R_\Box$ as a function of time at six different temperatures.
Inset: rate of change of $R_\Box$ as a function of temperature. The
line is a fit for $\alpha e^{-\frac{E}{kT}}$} \label{RvsT}
\end{figure}

Our samples are commercially available \cite{company} one sided
polished ${\rm SrTiO_3}$ crystals $(5 \times 5  \times 0.5 \ mm^3)$.
The ${\rm SrTiO_3}$ samples were irradiated  with Ar$^+$ ions,
accelerated with 4 kV and the beam's fluence was about $10^{15}$
ions per second per ${\rm cm^2}$. The estimated penetration depth of
the ions, L, in ${\AA}$ is given by the empirical formula
\cite{highly conductive,depth1}
$L=1.1\frac{E^{2/3}W}{\rho(Z^{1/4}_i+Z^{1/4}_t)^2}$ where E is the
energy in eV, W is the atomic weight of the target in atomic mass
units, $\rho$ is the target density, and $Z_i$, $Z_t$ are the atomic
numbers of the ions and the target, respectively (since ${\rm
SrTiO_3}$ is a compound, we  use for the target the weighted average
of the atomic weights and numbers). In our case $L\approx120{\AA}$;
therefore, we expect that the thickness of the conducting layer will
be on this order.

The samples become conducting when the irradiation time exceeds 30
sec and no more changes in conductivity are observed after several
minutes of irradiation. To irradiate specific parts of a substrate
in shapes that will allow resistivity and Hall measurement, we use
conventional photolithography that leaves 1 micron-thick photoresist
on the samples except for windows in the desired shapes.


Figure \ref{RvsT}a shows sheet resistance of Ar$^+$-irradiated ${\rm
SrTiO_3}$, determined with four point measurements. At low
temperatures a quadratic behavior is observed for the sheet
resistance ($R_\Box$), $R_\Box=a+bT^2$, typical for
electron-electron interactions (except for samples with very high
sheet resistance that exhibit resistivity minima). We note that the
residual resistivity ratio (RRR) in our samples exceeds in some
cases 500. Similar and even higher values of RRR have been reported
for electron-doped ${\rm SrTiO_3}$ and ${\rm SrTiO_3-LaAlO_3}$
heterostructures \cite{origin perspectives}.

The sheet resistance decreases significantly with irradiation until
saturation is obtained. As we can see from the inset of Figure
$\ref{RvsT}a$, there is a linear relation between sheet resistances
measured after different doses of irradiation (except for the range
of temperatures with resistivity minima - if it exists). This
indicates that in this range of doping there is no qualitative
change in the resistivity.

Time-dependent measurements (Figure $\ref{RvsT}b$) show that the
sheet resistance of the irradiated sample changes with time at a
temperature-dependent rate. Similar to the relation between sheet
resistance with different doses of irradiation, we find a linear
relation also between sheet resistances measured after different
waiting times  (see inset of Figure $\ref{RvsT}$a); namely, there is
no qualitative change in the resistivity behavior.

To extract the relevant energy scale for the relaxation in the sheet
resistance, we explore the temperature dependence of the relaxation
rate. As seen in the inset of Figure $\ref{RvsT}$b, this rate is
well fitted with Arrhenius law, $\alpha e^{-E/kT}$, where
E$\approx0.97 eV$. This activation energy is practically identical
to the activation energy of oxygen vacancies in $\rm SrTiO_3$
\cite{diffusion1} and $\rm Ba_{0.5}Sr_{0.5}TiO_3$ \cite{diffusion2}.
Hence, the observed relaxation is very likely due to diffusion of
oxygen vacancies. While we do not address here the change in the
conducting regions due to this diffusion, we note that irradiated
regions which are few microns apart remain electrically
disconnected.

In the following we explore how this diffusion affects other
transport properties.

\begin{figure}
\includegraphics[scale=0.5]{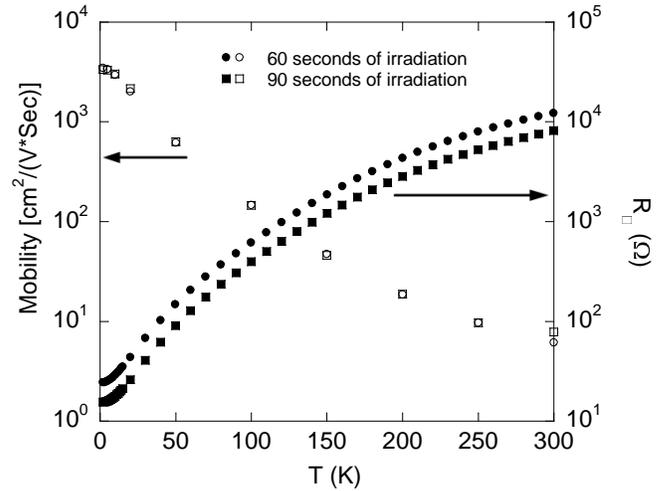}
\caption{Mobility and sheet resistance ($R_\Box$) in an Ar$^+$
irradiated ${\rm SrTiO_3}$ as a function of temperature after 60
(circles) and after 90 (squares) seconds of irradiation.}
\label{mobility}
\end{figure}

\begin{figure}
\begin{center}
\includegraphics [scale=0.55]{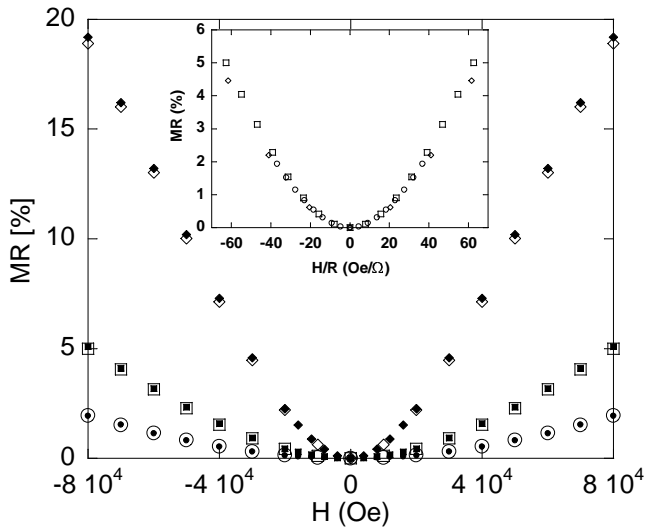}
\end{center}
\caption{Magnetoresistance of Ar$^+$ irradiated ${\rm SrTiO_3}$ at
50K (diamonds) 80K (squares) and 100K (circles) measured after two
different relaxation times (full and empty symbols). Inset: Scaling
of MR data (with a particular relaxation time) according to Kohler's
rule.} \label{MRvsHplot5}
\end{figure}

Figure \ref{mobility} shows the mobility and sheet resistance of one
of our irradiated samples after 60 and 90 seconds of irradiation. In
contrast to the resistance, the change of the mobility with
irradiation dose and relaxation time is hardly detectable. The
observed mobility is consistent in magnitude and temperature
dependence with previous reports \cite{origin perspectives}.

Figure \ref{MRvsHplot5} shows the MR at various temperatures
measured after two different relaxation times. Similar to the
mobility, the magnetoresistance $\Delta\rho/\rho$ does not change
with relaxation time or irradiation dose (not shown here) despite
the significant change in resistance. The inset of Figure
\ref{MRvsHplot5} shows that the MR data at temperatures higher than
50K obey Kohler's rule \cite{kohler}; namely, $\Delta\rho/\rho$
scales with $H/\rho$, implying it is a function of $H\tau$ alone
(where H is the magnetic field and $\tau$ is the scattering time).
That the  MR does not change between different relaxation times or
different irradiation times suggests that the scattering time is
practically unchanged.

In passing, we also note that, the MR bellow 50K does not obey
Kohler's rule and the angular dependance is different, indicating
that the mechanism of the MR at low temperatures is not the same as
above 50K.

The linear relation between sheet resistances with different doses
of irradiation and the linear relation between sheet resistances
measured after different waiting times, indicate that the
qualitative behavior of the resistivity does not change in a
detectable way. Together with the fact that the mobility and the
scattering time do not change when the sheet resistance changes, it
may suggest that the diffusion of the oxygen vacancies decreases the
number of charge carriers while hardly affecting the scattering rate
of the remaining charge carriers. It remains to be checked how this
observation is correlated with time dependent variation in the
thickness of the conducting layer and the spatial variation of
charge carriers density within this layer. Answers to these
questions are important for the potential use of electron-doped $\rm
SrTiO_3$ in future oxide electronics.

L.K. acknowledges support by the Israel Science Foundation founded
by the Israel Academy of Science and Humanities.

\end{document}